\documentclass[a4paper,twoside]{article}

\usepackage{epsfig}
\usepackage{subcaption}
\usepackage{calc}
\usepackage{amssymb}
\usepackage{amstext}
\usepackage{amsmath}
\usepackage{amsthm}
\usepackage{multicol}
\usepackage{mathptmx}
\usepackage{apalike}
\usepackage{CJKutf8}
\usepackage{geometry}
\usepackage{SCITEPRESS}     

\usepackage{url}
\PassOptionsToPackage{hyphens}{url}

\DeclareMathOperator*{\argmin}{arg\,min}

\begin{document}

\title{The Complexity of Social Media Response: \\ Statistical Evidence For One-Dimensional Engagement Signal in Twitter}

\author{\authorname{Damian Konrad Kowalczyk\sup{1,2}\orcidAuthor{0000-0002-5612-0859}, Lars Kai Hansen\sup{2}\orcidAuthor{0000-0003-0442-5877}}
\affiliation{\sup{1}Microsoft Development Center Copenhagen, Business Applications Group, Kanalvej 7}
\affiliation{\sup{2}Technical University of Denmark, Department of Applied Mathematics and Computer Science, Matematiktorvet 303B}
\affiliation{2800 Kongens Lyngby, Denmark}
\email{dakowalc@microsoft.com, \{damk, lkai\}@dtu.dk}
}


\abstract{Many years after online social networks exceeded our collective attention, social influence is still built on attention capital. Quality is not a prerequisite for viral spreading, yet large diffusion cascades remain the hallmark of a social influencer. Consequently, our exposure to low-quality content and questionable influence is expected to increase. Since the conception of influence maximization frameworks, multiple content performance metrics became available, albeit raising the complexity of influence analysis. In this paper, we examine and consolidate a diverse set of content engagement metrics. The correlations discovered lead us to propose a new, more holistic, one-dimensional engagement signal. We then show it is more predictable than any individual influence predictors previously investigated. Our proposed model achieves strong engagement ranking performance and is the first to explain half of the variance with features available early. We share the detailed numerical workflow to compute the new compound engagement signal. The model is immediately applicable to social media monitoring, influencer identification, campaign engagement forecasting, and curating user feeds.}

\onecolumn \maketitle \normalsize \setcounter{footnote}{0} \vfill


\section{Social media engagement}
The unprecedented amount of attention aggregated by online social networks  comes under intense criticism in the recent years \cite{Bueno2016,Wu2017,Beyersdorf2019,Bybee2019}, as billions are now exposed to low-quality content and questionable influence. Platforms like Facebook and Twitter, offer an unparalleled opportunity for influence analysis and maximization, impacting public opinion, culture, policy, and commerce \cite{davenport2001attention}.

Extant work on influence analysis focuses on homogeneous information networks and attributes the greatest influence to authors triggering the largest diffusion cascades \cite{Franck2019}. When the author's influence is modeled as the ability to maximize the expected spread of information in the network \cite{Pezzoni2013,Soheil2019}, the most desirable user-generated content is the one propagated furthest, in Twitter measured by the number of retweets. Propagation metrics however (retweet count in particular), do not capture the average individual attention received. Retweet action does not inform, e.g., if the actor has actually read the content, let alone consider the source or whether that effort was left to the followers. Meanwhile, the abundance of information to which we are exposed through online social networks is exceeding our capacity to consume it \cite{Weng2012}, let alone in a critical way. Work presented in \cite{Weng2012,Qiu2017} shows that content quality is not a prerequisite for viral spreading, and \cite{Lorenz-Spreen2019} shows that the competition for our attention is growing, causing individual topics to receive even shorter intervals of collective attention. Accordingly, our exposure to low-quality information and, by extension low-quality influence is increasing (Table \ref{table:tweets}).
\begin{table}[ht]
\centering
\caption{Four popular tweets ranked by the most prevalent influence predictor: size of diffusion triggered in the network, in Twitter measured by the number of retweets}
\resizebox{\columnwidth}{!}{%
\begin{tabular}{|l|c|c|c|}
\hline
\textbf{Tweet (body)}         
& \textbf{Retweets} & \textbf{Replies} & \textbf{Favorites} \\ \hline
\begin{CJK*}{UTF8}{gbsn}
"ZOZOTOWN新春セールが史上最速で取扱高100億円を先ほ(...)"\end{CJK*} & \textbf{4.5M}          & 357.4K             & 1.3M              \\ \hline
"HELP ME PLEASE. A MAN NEEDS HIS NUGGS"                                                                                                                     & \textbf{3.47M}             & 37K              & 0.99M              \\ \hline
"If only Bradley's arm was longer. Best photo ever. \#oscars"                                                                                               & \textbf{3.21M}             & 215K             & 2.29M              \\ \hline
\begin{tabular}[c]{@{}l@{}}"No one is born hating another person because of the color \\of his skin or his background or his religion..."\end{tabular} & \textbf{1.61M}             & 69K              & 4.44M              \\ \hline
\end{tabular}%
}
\label{table:tweets}
\end{table}
%
Today, the digital footprint of an audience goes far beyond the retweet action. Platforms like Facebook and Twitter record an increasingly diverse set of user behaviors, including number of clicks, replies or favorites (likes). 
Since the work of \cite{Pezzoni2013}, Twitter has made many of these metrics available to the public, inviting a more holistic approach to influence modeling, albeit rising the complexity of all dependent tasks. Consequently, few studies to date systematically investigate how to model the strength of influence in heterogeneous information networks, and the processes that drive popularity in our limited-attention world remain mostly unexplored \cite{Franck2019,Weng2012}.
\newline \indent The four Tweets in Table \ref{table:tweets} illustrate that the mechanisms leading to high engagement are complex.  In the following work, we investigate the multi-dimensional response of on-line audiences to understand this complexity. 
We examine and consolidate multiple discrete engagement metrics towards a new compound engagement signal. While the new signal is statistically motivated, we next show the relevance of the signal for understanding engagement in multiple datasets. In particular, we show that the new signal is more predictable than the individual metrics (e.g., diffusion size measured by retweet count) prevalent in literature. Our engagement model is the first to explain half of the variance with features available early, and to offer strong \cite{cohen1988spa} ranking performance simultaneously. We provide the workflow for calculating the new compound engagement signal from the raw count.\\

The contributions of this paper are summarized as follows:

\begin{enumerate}
    \item Parallel analysis of three individual content performance signals, showing evidence of one-dimensional engagement signal on Twitter
    \item new compound engagement formula, capturing over 75\% of variance in available engagement signals
    \item advancing feature representation of user generated content on Twitter, to consider increasingly popular 'quote tweets', validated on two real-world datasets
    \item two new engagement models (response and popularity), delivering strong ranking performance
    \item new state-of-the-art in virality prediction on Twitter
    \item finally, a new more holistic, compound engagement model, first to explain half of the variance with content features available at the time of posting, and to offer strong ranking performance simultaneously
\end{enumerate}

\section{Methodology}

In this section we describe the application of unsupervised learning towards contributions (1,2,6), data collection and feature extraction approach towards contribution (1,3), and the chosen supervised method towards contributions (4,5,6). 

\subsection{Principal Engagement Component}
We acquire the multivariate set of responses forming the ground truth vector:
\begin{equation}
    e_{gt} = [e_{\rm retweets}, e_{\rm replies}, e_{\rm favorites}]^T.
    \label{equation_e}
\end{equation}
Recent work on engagement modeling, e.g.,  \cite{lee2018advertising} defines any response as a sign of engagement, effectively reducing the multivariate response to a one-dimensional signal. However, to our knowledge, the complexity of the engagement signal has not been explored more formally.  While it appears credible that the population response signals,i.e., the dimensions of the of vector ${\bf e}$, are highly correlated, we can test the effective dimension of the space populated by the vectors using so-called Parallel Analysis (PA)
\cite{horn1965rationale,jorgensen2011model}. In PA principal component analysis of the measured signals is compared with the distribution of the principal components of null data obtained by permutation under a (null) hypothesis that there is no dependency between the individual response signals. Consistent with this hypothesis, we can permute the sequence of the signals for each observation separately. In particular, we compute the upper $95\%$quantile for the distribution of the eigenvalues in the permuted data. 
Eigenvalues of the original unpermuted data set that reject the null hypothesis are considered "signal".

Principal components are computed on the response signals subject to a variance stabilization transformation,
\begin{equation}
    e=\ln(e_{gt}+1),
    \label{equation_variance}
\end{equation}
see e.g., \cite{Can2013,Kowalczyk2018}. 

\subsection{Projection on the engagement component}
Hypothesizing a one-dimensional engagement signal, we compute the value as the projection on the first principal component of the transformed data of dimension $D=3$,
\begin{equation}
    \rm{E}_1=\sum_{i=1}^{D}w_i \left( \ln(e_i+1) - \mu_i \right),
    \label{equation_projection}
\end{equation}
where $\mu_i = \frac{1}{N}\sum_{n=1}^N\hat{e}_{i,n}$ is the $i$'th component of the $D$-dimensional mean vector for a sample of size $N$, while $w_i$ is the $i$'th component of the first principal component, computed on the same sample.

\subsection{Gradient Boosted Regression Trees (GBRT)}
We consider the problem of predicting audience engagement for a given tweet based on features available immediately after its delivery (Table 3). Features describing the author are used together with the content, language, and temporal descriptors to predict the size of retweet cascade, number of likes, number of replies, and the proposed compound engagement signal. GBRT is a tree ensemble algorithm that builds one regression tree at a time by fitting the residual of the trees that preceded it. The training process minimizing a chosen twice-differentiable loss function can be described as
\begin{equation}
    \theta^*=\argmin_\theta \sum_{i=1}^{N}L_{\rm{SE}}(\hat{e}_{i}, e_i),
\end{equation}
where \(\theta\) contains all parameters of the proposed model, N is the number of examples, and \(L_{\rm{SE}}\) is the squared error of an individual prediction,
\begin{equation}
    L_{\rm{SE}}(e, \hat{e})={(e - \hat{e})^2}.
\end{equation}
We follow \cite{Can2013,Kowalczyk2018} to stabilize variance of all individual engagement signals via log-transformation as in Equation \ref{equation_variance}.

\subsubsection{Gradient Boosting Framework} 
We use Microsoft's implementation of Gradient Boosted Decision Trees \cite{Ke2017} for model training and tuning. LightGBM offers accurate handling of categorical features by applying \cite{WalterD.Fisher1958}, which limits the dimensionality of our tasks. 

\section{Data Collection}
Recent work on social network analysis re-emphasizes the importance of dataset size, to make reliable predictions from representative samples. The larger the dataset, the better the accuracy and consistency of a predictive model because it minimizes the possibility of bias. However, as argued by \cite{Agarwal2019}, this intuition is incomplete. Relying solely on short timeframe samples or keyword-based crawling can produce a large dataset full of noise and irrelevant \cite{Bhattacharya2017} data. Careful collection and filtering strategies, in addition to large-scale sampling, are critical for building datasets representative of the population and engagement modeling at scale.

\subsection{Unique Tweets}
\begin{figure}[h] 
\includegraphics[width=\columnwidth]{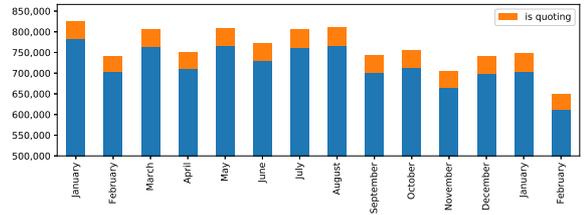}
\caption{T2017-ML volume per month: Historical APIs allow near uniform sampling of large-scale data to ensure higher proportion of the population in a sample}
\label{fig:volumePerMonth}
\end{figure}

We use Twitter Historical PowerTrack APIs to collect training and validation datasets described in Table \ref{table:datasets}. Retroactive filtering of Twitter archives allows close reproduction of datasets used in prior work (where still public) e.g., \cite{Wang2018a,Kowalczyk2018}. Historical PowerTrack API also enables near-uniform sampling across long time-frames (Figure \ref{fig:volumePerMonth}), to increase the proportion of the population in a sample, as motivated by \cite{Kim2018}. Collecting a dataset similar to T2017-ML by sampling Twitter Firehose prevalent in prior work, would have taken 14 months. 

\subsection{Engagement totals}
Three content engagement metrics are made publicly available by Twitter since 2015. We use Twitter's Engagement Totals API to retrieve the number of retweets, replies, and favorites ever registered for each tweet (even if removed later via unlike or account suspension). Use of the Engagement Totals API ensures 100\% accuracy of our supervisory vector of response signals $e$.
%

\subsection{Sentiment prediction}
\cite{Hansen2011c,Kowalczyk2018} show the impact of sentiment on tweet's virality (retweetability). We reuse sentiment predictions from \cite{Kowalczyk2018} for all tweets in the validation datasets to explore correlation with other engagement metrics and ensure fair comparison with previous results. The analysis was performed for tweets in 18 languages, using Text Analytics APIs from Microsoft Cognitive Services \cite{TextAnalytics}.

\begin{table}[h]
\centering
\caption{Datasets acquired}
\resizebox{\columnwidth}{!}{%
\begin{tabular}{l|c|c|c} \hline
\textbf{Dataset}       & \textbf{T2016-IMG}       & \textbf{T2017-ML}   & \textbf{T2018-ML}       \\ 
\hline
introduced & Wang (2018) & Kowalczyk (2018) & now \\ 
w/image only & True            & False  & False         \\ 
languages     & \textbf{English}     & \textbf{18}   & \textbf{all}           \\ 
months total   & \textbf{3}               & \textbf{14}  & \textbf{12}              \\ 
month from     & 2016.10 & 2017.01 & 2018.01 \\ \hline
unique tweets & 2,848,892       & 9,719,264  & 29,883,324     \\ 
quoting       & 421,175                & 583,514 & 2,647,072               \\
retweets total      &  5,929,850               & 11,361,699 & 42,919,158           \\ 
replies total      &  717,644               & 3,576,976   & 12,414,907           \\ 
favorites total    & 12,665,657                & 29,138,707  & 134,523,998           \\ 
no engagement      & 1,547,829                & 5,689,501  & 14,813,772            \\ 
\hline
\end{tabular}
}
\label{table:datasets}
\end{table}

\subsection{Datasets}
Table \ref{table:datasets} offers a summary of three datasets collected for this study.

\begin{enumerate}
	\item \textbf{T2016-IMG} to evaluate our feature representation and training method in  comparison with the work of \cite{Mazloom2016,McParlane2014,Khosla2014,Cappallo2015,Wang2018a,Kowalczyk2018}. The dataset matches the same filters, as applied before (timeframe, language code or the presence of an image attachment).
	\item \textbf{T2017-ML} to evaluate the generalizability of our resulting models across seasons and languages (cultures) and  comparison with the work of \cite{Kowalczyk2018}. This dataset represents a near-uniform sample of Twitter 2017 volume in all 18 languages supported by the sentiment analysis service \cite{TextAnalytics}.
	\item \textbf{T2018-ML} to evaluate the generalizability of our compound engagement signal across years. This dataset represents a near-uniform sample of entire Twitter 2018 volume in all known languages. In this study, T2018-ML dataset is used in unsupervised experiments only.
\end{enumerate}

\noindent Datasets T2016-IMG and T2017-ML are split into 70\% training, 20\% test and 10\% validation sets. To aid reproducibility, we share unique ID's of acquired tweets along with sentiment predictions.

\subsubsection{Privacy respecting storage}
The data analyzed in this study is publicly available during collection. How much of it remains public, can change rapidly afterward. We follow the architecture proposed by \cite{Kowalczyk2018} to secure the data in a central highly scalable database, exposed to applicable privacy requests from Twitter’s Compliance Firehose API, and to feature extraction requests from our Spark cluster.
\begin{table}[h]
\caption{Feature representation summary}
\resizebox{\columnwidth}{!}{%
\begin{tabular}{lccc}
\hline
\textbf{Feature}        & \textbf{Representation} & \textbf{Skewness} & \textbf{Quoted\textsuperscript{\textdagger}} \\ \hline
\textbf{followers count}         & ordinal       & 0.212             & \textbf{True}            \\
\textbf{friends count}           & ordinal       & -0.321            & \textbf{True}            \\
\textbf{account age (days)}      & ordinal       & 0.203             & \textbf{True}            \\
\textbf{statuses count}          & ordinal       & -0.665            & \textbf{True}            \\
\textbf{actor favorites count}   & ordinal       & -1.023            & \textbf{True}            \\
\textbf{actor listed count}      & ordinal       & 0.687             & \textbf{True}            \\
\textbf{actor verified}          & categorical   & -                 & \textbf{True}            \\ \hline
\textbf{body length}             & ordinal       & -1.426            & \textbf{True}            \\
\textbf{mention count}           & ordinal       & 3.820             & \textbf{True}            \\
\textbf{hashtag count}           & ordinal       & 5.808             & \textbf{True}            \\
\textbf{media count}             & ordinal       & 3.203             & \textbf{True}            \\
\textbf{url count}               & ordinal       & 1.449             & \textbf{True}            \\
\textbf{language code}           & categorical   & -                 & \textbf{True}            \\
sentiment value         & continuous    & -0.014            & False           \\ \hline
posted hour             & ordinal       & -0.058            & False           \\
posted day              & ordinal       & 0.021             & False           \\
posted month            & ordinal       & 0.210             & False           \\ \hline\hline
retweet count           & label         & 6.091             & n/a             \\
reply count             & label         & 2.330             & n/a             \\
\textbf{favorite count}          & label         & 3.122             & \textbf{True}            \\ \hline
\end{tabular}
}
\small\textsuperscript{\textdagger} if True, additional feature is extracted from the quoted tweet
\label{table:features}
\end{table}

\subsubsection{Feature extraction}

 Table 3 describes features extracted from each tweet. To ensure scalability in production, only the information available at the time of engagement is considered.
 In 2015 Twitter introduced ‘quote retweets’ (or ‘quote RTs’) impacting political discourse and its diffusion as shown by \cite{Garimella:2016:QRT:2908131.2908170}. Over 3.5 million tweets collected for this study quote another (Table \ref{table:features}). We extend the feature representation by \cite{Kowalczyk2018} to represent them. Table 3 shows in bold, an additional 14 unique features computed for quoted RT's. We log-transform highly skewed (count of followers, friends, statuses, and number of times the actor has been listed) to stabilize variance.

\section{Results}

We begin with examining all available content performance signals (count of retweets, replies and favorites) in the extended time-frame datasets. We look for potential correlations that could enable reducing the dimension of engagement using Parallel Analysis. In the supervised experiments, first we evaluate our methodology and feature representation against previous state-of-the-art methods, by modelling the individual influence metrics (e.g. virality) and the compound engagement on the benchmark dataset T2016-IMG. Finally we evaluate the generalizability of our method across topics and cultures, modeling engagement on the multilingual extended-timeframe dataset T2017-ML.

\subsection{Evidence for a one-dimensional engagement signal}

\begin{figure}[t!]
\includegraphics[width=\columnwidth]{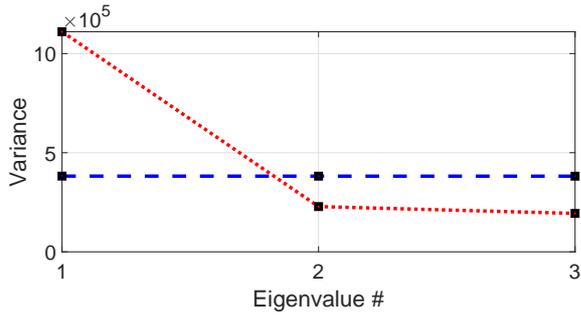}
\caption{Parallel Analyses of the response signals for the 2017 data set provide evidence for a one-dimensional engagement signal: Only the first component ('$1$'- red dotted line) exceeds the $95\% $ quantile of the corresponding eigenvalue  in the null hypothesis (blue dashed line).}
\label{fig:PA}
\end{figure}

We perform Parallel Analysis and compute the principal components and their associated projected variances for the log-transformed data as well as for $Q=100$ permutations of the data assuming the no correlation null. The one-sided upper $95\%$ quantile is computed from the permuted samples. Variances of the un-permuted signals and the $95\%$ quantiles for the three eigenvalues of the permuted data are shown in figure \ref{fig:PA}.
Very similar results  are obtained for the 2018 data set (not shown).

\subsection{The engagement signal}
We perform principal component analysis of the two data sets keeping a single principal component. The mean vectors and projections are found in Table \ref{table:principal}. The variance explained by the first components in the three analyses: $2016: 83\%, 2017: 72\%, 2018: 77\%$.
%
%
%
\begin{table}[h]
\caption{First principal components of the extended time-frame engagement signals,  used to compute the one-dimensional compound engagement (see Equation \ref{equation_projection})}
\resizebox{\columnwidth}{!}{%
\begin{tabular}{lcccccc}
                              & \multicolumn{2}{c}{\textbf{retweets}}                            & \multicolumn{2}{c}{\textbf{replies}}                             & \multicolumn{2}{c}{\textbf{favorites}}      \\ \cline{2-7} 
\multicolumn{1}{l|}{}         & \multicolumn{1}{c|}{\(w_1\)}    & \multicolumn{1}{c|}{\(\mu_1\)}    & \multicolumn{1}{c|}{\(w_2\)}    & \multicolumn{1}{c|}{\(\mu_2\)}    & \multicolumn{1}{c|}{\(w_3\)}    & {\(\mu_3\)}    \\ \hline
\multicolumn{1}{l|}{\textbf{T2017-ML}} & \multicolumn{1}{c|}{0.451} & \multicolumn{1}{c|}{0.049} & \multicolumn{1}{c|}{0.145} & \multicolumn{1}{c|}{0.082} & \multicolumn{1}{c|}{0.880} & {0.148} \\ \hline
\multicolumn{1}{l|}{\textbf{T2018-ML}} & \multicolumn{1}{c|}{0.450} & \multicolumn{1}{c|}{0.066} & \multicolumn{1}{c|}{0.188} & \multicolumn{1}{c|}{0.080} & \multicolumn{1}{c|}{0.872} & {0.205} \\ \hline
\end{tabular}
}
\label{table:principal}
\end{table}
\subsection{Predicting Engagement}
\textbf{Metrics} We compute the Spearman \(\rho\) ranking coefficients to measure each model's ability to rank the content depending on the definition of engagement. We compute the relative measure of fit \(R^2\) to compare the variance explained in the compound engagement and in the individual engagement signals. The absolute measure of fit (RMSE) is chosen as an objective of optimization, to penalize large errors and relative insensitivity to outliers. The \(p\)-value for all reported \(\rho\) results is \(p<0.001\). Each metric is an average from 3-fold cross-validation. SciPy version 1.3.1 is used to ensure \(\rho\) tie handling. Interpretation of \(R^2\) and Spearman \(\rho\) is domain-specific, with guidelines for social and behavioral sciences proposed by \cite{cohen1988spa}.\\

\noindent\textbf{Representation} First round of our supervised experiments focus on evaluating our user-generated content feature representation and GBRT approach against previous state-of-the-art methods, in modeling established engagement signals, like the size of diffusion (e.g., retweet count), response (i.e., number of replies) and popularity (i.e., number of favorites/likes), before attempting to predict the compound engagement. Table \ref{benchmark} shows the performance of our GBRT with RMSE objective and new feature representation. Features extracted from the quoted content did not provide a significant boost over SOTA, likely due to visual modality dominating in the T2016-IMG dataset, as considered by \cite{Wang2018a}. The approach did, however, match the performance of \cite{Kowalczyk2018} in virality ranking, and achieves strong \cite{cohen1988spa} performance without considering image modality. Applied to predict the new compound engagement, it sets a new benchmark for content engagement ranking \(\rho=0.680\).
%
%
\begin{table}[h]
\caption{Method evaluation on the T2016-IMG dataset.}
\resizebox{\columnwidth}{!}{%
\begin{tabular}{l|c|c|c}
\hline
\textbf{Method}     & $R^2$ & \(\rho\) & RMSE \\ \hline
\cite{McParlane2014}\textsuperscript{\textdagger}  & -                             & 0.257        & -                   \\
\cite{Khosla2014}\textsuperscript{\textdagger}     & -                             & 0.254        & -                     \\
\cite{Cappallo2015}\textsuperscript{\textdagger}   & -                             & 0.258        & -                   \\
\cite{Mazloom2016}\textsuperscript{\textdagger}    & -                             & 0.262        & -                  \\
\cite{Wang2018a}\textsuperscript{\textdagger}       & -                             & 0.350        & -                    \\
\cite{Kowalczyk2018}  & 0.391                         & 0.504       & 0.555          \\ \hline
\textbf{virality} (retweets)   &  0.393             &  0.504     & 0.554           \\
\textbf{response} (replies)  & 0.239        &  0.384         &  0.290             \\
\textbf{popularity} (favorites) & 0.500        &  0.656      &  0.665            \\
\hline
\textbf{engagement} (compound) &  \textbf{0.501}    & \textbf{0.680}    & \textbf{0.341}           \\
\hline
\end{tabular}%
}
\small\textsuperscript{\textdagger} independent evaluation by \cite{Wang2018a}
\label{benchmark}
\end{table}

\noindent \textbf{Engagement} The second round of supervised experiments focuses on the scalability and generalizability of our approach across topics and cultures (languages). Table \ref{multiexp} shows the performance of our engagement models on the multilingual extended timeframe dataset. Predicting the number of retweets with our new feature representation outperforms \cite{Kowalczyk2018}, offering new state-of-the-art in virality ranking. Response and popularity models achieve strong \cite{cohen1988spa} ranking performance on T2017-ML. The compound engagement model again shows an increase in ranking performance over all individual engagement models, setting a new benchmark for engagement variance explained \(R^2=0.507\). Table \ref{table:tweets2} offers a real-world illustration of Engagement ranking performance, radically different than traditional diffusion-based ranking exemplified in Table \ref{table:tweets}.
\begin{table}[t]
\caption{Engagement prediction performance on T2017-ML dataset. 
\(SD < 0.001\) across 3-fold CV}
\resizebox{\columnwidth}{!}{%
\begin{tabular}{|l|c|c|c|c|c}
\hline
\textbf{Method}     & $R^2$ & $\rho$  & RMSE \\ \hline
\cite{Kowalczyk2018}  & 0.402   & 0.369     & 0.336                \\ \hline
\textbf{virality} (retweets)  & 0.425      &   0.371   & 0.329     \\
\textbf{response} (replies)   &  0.302   & 0.512       & 0.292      \\
\textbf{popularity} (favorites)   & 0.493      & 0.526        & 0.484 \\ \hline
\textbf{engagement} (compound) & \textbf{0.507}    & \textbf{0.529}    &  \textbf{0.228}   \\\hline
\end{tabular}%
}
\label{multiexp}
\end{table}
%
%
\begin{figure}
\includegraphics[width=\columnwidth]{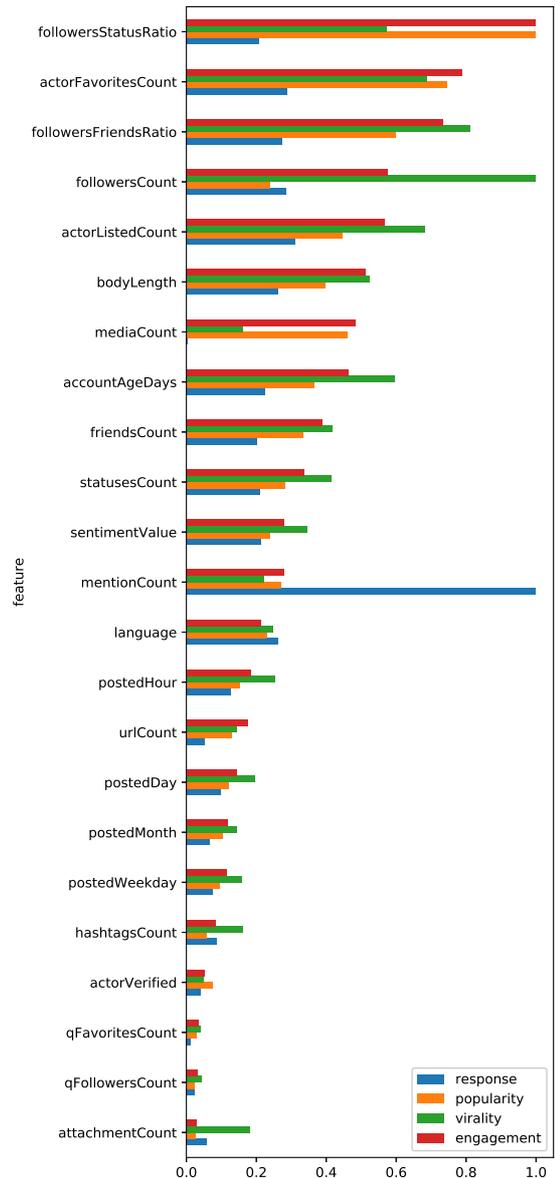}
\caption{Relative feature importance depending on the definition of engagement (top 23 out of 31 features).}
\label{fig:featureImp}
\end{figure}
\begin{table}[h]
\caption{Four popular tweets, ranked by the new compound engagement metric}
\resizebox{\columnwidth}{!}{%
\begin{tabular}{|l|c|c|c}
\hline
\textbf{Tweet (body)} & \textbf{Engagement} \\ \hline
\begin{tabular}[c]{@{}l@{}}"No one is born hating another person because of the color \\  of his skin or his background or his religion..."\end{tabular} & \textbf{9.283}    \\ \hline
"If only Bradley's arm was longer. Best photo ever. \#oscars" &  \textbf{9.266}             \\ \hline
\begin{CJK*}{UTF8}{gbsn}
"ZOZOTOWN新春セールが史上最速で取扱高100億円を先ほ(...)"\end{CJK*}  & \textbf{9.158}              \\ \hline
"HELP ME PLEASE. A MAN NEEDS HIS NUGGS"  &   \textbf{8.822}            \\ \hline
\end{tabular}%
}
\label{table:tweets2}
\end{table}
\subsection{Feature Importance}
Figure \ref{fig:featureImp} offers a comparison of feature importance between all engagement models trained on the T2017-ML dataset. The importance equals total gains of splits which use the feature, averaged across 3-folds and rescaled to \([0, 1]\) for comparison across all engagement models. The uncertainty for virality features does not exceed 6\%. When predicting response (i.e., number of replies), we find the number of users mentioned to have the highest predictive value, while the number of image attachments (i.e., media count) to have almost none. The number of followers, most popular in all prior work on virality prediction is fourth when predicting compound engagement. The average number of followers received with each status or number of times the author liked another tweet is far more predictive of compound engagement.
\section{Conclusion}
In this study, we have analyzed the complexity of the multivariate response of users engaging with social media. We have employed large-timeframe collection and filtering strategies to build datasets of unique tweets that could better represent Twitter's population. We have acquired, examined, and consolidated various response (engagement) metrics available for each of the tweets. The significant correlation found between individual response signals leads us to propose a new one-dimensional compound engagement signal. We showed on multiple benchmark datasets, that compound engagement is more predictable than any individual engagement signal, most notably the number of retweets, measuring the size of diffusion cascade, predominant in influence maximization frameworks. \cite{Franck2019,Soheil2019}. 

Our compound engagement model is the first to explain half of the variance with features available at the time of posting, and to offer strong \cite{cohen1988spa} ranking performance simultaneously. The model is ready for production with immediate application to social media monitoring, campaign engagement forecasting, influence prediction, and maximization. We propose the ability to engage the audience as a new, more holistic baseline for social influence analysis. We share the compound engagement workflow and parameters (Eq.\ (\ref{equation_projection}) and Table (\ref{table:principal})) to ensure reproducibility and inspire future work on engagement modeling. We hope the future work will balance any negative impact of diffusion-based influence maximization, on our collective attention and well-being.

\subsection{Acknowledgements}
This project is supported by the Business Applications Group within Microsoft and the Danish Innovation Fund, Case No. 5189-00089B. We would like to acknowledge the invaluable support of Sandeep Aparajit, Jörg Derungs, Ralf Gautschi, Tomasz Janiczek, Charlotte Mark, Pushpraj Shukla, and Walter Sun. Any opinions, findings, conclusions, or recommendations expressed in this material are those of the authors and do not necessarily reflect those of the sponsors.

\bibliographystyle{apalike}
\bibliography{engagement.bib}
\end{document}